\documentclass[aps,prl,amsmath,amssymb,twocolumn]{revtex4-1}
\usepackage[utf8]{inputenc}
\usepackage{bm}        
\usepackage{verbatim}   
\usepackage{braket}
\usepackage{afterpage}
\usepackage{amsmath}
\usepackage{nth}
\usepackage{physics}
\usepackage{graphicx}

\newcommand{\im}{\ensuremath{ \text{i} }}

\newcommand{\Ef}{\ensuremath{ \mathbf{E}_f }}
\newcommand{\Eb}{\ensuremath{ \mathbf{E}_b }}
\newcommand{\di}{\ensuremath{ \mathbf{d} }}

\begin{document}

\title{Reflection by two level system: phase singularities on the Poincar{\'e} hypersphere}


\author{Ben Lang$^{1}$, Edmund Harbord$^{2}$ and Ruth Oulton$^{2}$}
\affiliation{$^{1}$George Green Institute for Electromagnetics Research, Faculty of Engineering, The University of Nottingham, NG7 2RD, UK\\
$^{2}$Quantum Engineering Technology Labs, H. H. Wills Physics Laboratory and Department of Electrical \& Electronic Engineering, University of Bristol, BS8 1FD, UK}

\begin{abstract}
We consider the reflection of a photon by a two-level system in a quasi-one-dimensional waveguide. This is important in part because it forms the backdrop for more complicated proposals where many emitters are coupled to the waveguide: leading to super and subradiant coupling even when the emitters are distant. The incorporation of chiral effects, for example unidirectional emission of dipole emitters, has already led to rich physics such as dimer coupling. However, chirality is not the only effect of the dipole, as we explore from a phase singularity perspective. We demonstrate that control of the dipole allows a rich variety of control of the phase and amplitude of the scattered light in both directions. This expands the scope for the physics of 1D chains of emitters.
\end{abstract}

\maketitle

The exploration of quantum emitter systems coupled to ``one-dimensional" waveguide-like photonic structures has developed into a wide field over the past few years. The modified photonic density of states allows near-perfect coupling between quantum emitters in the waveguide that does not decay with distance.  Studies of arrays of atoms or quantum emitters in such systems predict a rich variety of physics. For instance, carefully positioned emitters are predicted to show superradiance effects \cite{Gonzalez_2013_Entanglement}. Unidirectional emission and scattering leads to symmetry breaking in the coupling of arrays of atoms, and the formation of ``dimers" of many-body dark states \cite{Zoller_2014_dimer}. In those studies, one exploits the interference between back-scattered and forward-scattered light. However, these studies make a priori assumptions, for example that the phase change on reflection is always $\pi$ \cite{chiral_quantum_optics}, or only considering the specific case of forward scattering for chiral emitters at specific points in the waveguide.

In this work we explore generalised complex dipole systems in a waveguide with generalised polarization texture. The result is that one has signficant control over the phase and amplitude of the scattered light, particularly in reflection. Such generalized dipole properties will diversify the capabilites of one-dimensional atom-chain systems, and relax the positioning requirements for those emitters.

\begin{figure}[t]
\includegraphics[scale=0.38]{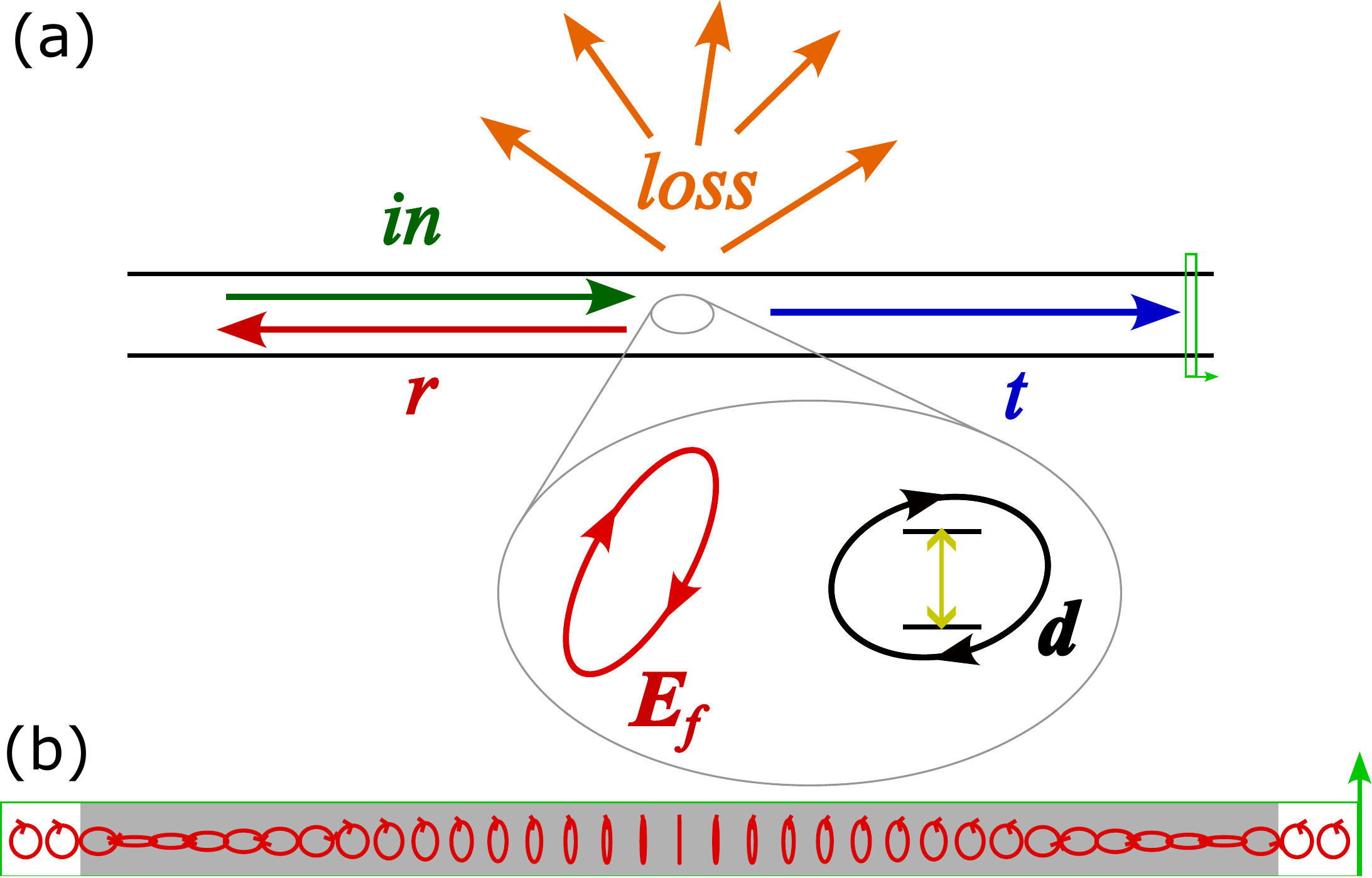}
\caption{(a) Schematic cartoon of the scattering process. The input is split between reflection, transmission and loss, with the coefficients determined by the waveguide electric field at the location of the TLS (red), and the transition dipole of the TLS (black). (b) Polarisation structure in a simple waveguide, consisting of a high-index rectangular block (grey) in air (white). Polarisation ellipses are shown through a cut, with light propagating up the page (frame arrow).}
\label{scattering_cartoon}
\end{figure}

For simplicity we consider the reflection of a photon from just one two level system (TLS), a classic problem in 1D quantum optics \cite{Fan_2005_reflection}. A photon incident on the TLS can be reflected, transmitted or scattered out of the waveguide, ie. lost (although in a 1D system these losses are assumed to be small). These possibilities are summarised using complex reflection and transmission coefficients, $r$, $t$, with $|r|^2 + |t|^2 \leq 1$ with equality occurring only at zero loss. The situation is depicted in fig.\ref{scattering_cartoon},(a).

The generalised nature of our study is that it includes both the waveguide polarisation at the TLS location (which, even for simple waveguide is very sensitive to location, fig.1(b)) and the electric dipole of the TLS. Recently there has been interest in so called  ``chiral" coupling, which exploits the longitudinal component of the waveguide electric field to make the TLS couple asymmetrically to the waveguide modes in either direction \cite{Rauschenbeutel_2014_scatter, Zayats_2013_surface_waves, Leuchs_2015_photonic_wheels, Coles_2016_chiral, Zayats_2018_Janus}. Chiral coupling enables systems where the reflection/transmission coefficients are different for light incident on the TLS from either side. For example a TLS decoupled from the forwards mode will have $t=+1$ for a photon injected forwards, but a photon incident in the backward direction may have $t$ anywhere between $+1$ and $-1$ depending on the coupling strengths to the backward and loss modes (the applications of $t=0$ are discussed in \cite{Alejandro_2016_diode, Rauschenbeutel_2015_diode} and $t=-1$ in \cite{Young_2015_chiral, lodahl_2015_chiral}).

The complex coefficients $r$ and $t$ depend on both the exact location in the waveguide and the properties of the dipole. In nanophotonic systems the electric E-field polarization at any specific point $r$ varies such that one can describe a polarization ellipse at each point $r$ \cite{Young_2015_chiral}. The ellipse gives the relative magnitudes of the $E_x$ and $E_y$ components of the field and their relative phase. It is common to see purely linear E-field points, purely circular (C) points, where the $E_x$ and $E_y$ components are equal and out of phase by $\pi /2$, and arbitrary elliptical points in between those two extremes. Unidirectional chiral coupling occurs when a circular dipole emitter (ie a linear combination of $d_x$ and $d_y$ dipole out of phase by $\pi /2$) is placed at a C-point (the direction of emission is dictated by the sign of the phase between the dipole components). However one can construct the arbitrary complex dipole $\mathbf{d} = \alpha d_x + \beta d_y$. Our previous work \cite{Lang_2022_perfect}, has highlighted that even when emitters are placed at an elliptical point, emission may be made unidirectional by matching the helicity and eccentricity of the ellipse and the dipole, but putting the long axes of the two ellipses orthogonal. This negates the backscattered component so that only forward scattering can occur.

We now explore the case where reflection is desired, by considering an arbitrary dipole where the linear component is able to couple to the backscattered direction. What is interesting here is the phase of the reflection. While input-output models give only a $\pi$ phase shift to the backscattered light, we show here that the orientation of the ellipse governs the phase of the reflected light, giving rise to a rich structure. This is an extremely useful property as it allows one to tune the phase delay when considering a chain of atoms. This means that one is no longer constrained to precise positions of the emitters in a 1D chain. Also, by dynamically controlling the dipole ellipse orientation, one may control the reflectivity in-situ.


We consider electric dipole interactions between the TLS and the waveguide, and assume a narrow-band photon (narrow in frequency, long in time). The crucial parameters determining the single photon $t$, and $r$ coefficients are the local polarisation of the forwards (backwards) waveguide mode at the TLS position $\mathbf{E}_{f(b)}$ and the electric dipole of the TLS transition $\di$. Both $\Ef$ and $\di$ are complex vectors in space. Without loss of generality we assume a basis in which both are 2D. The time-reversal symmetry of Maxwell's equations requires that $\Ef = \Eb^*$, which amounts to reversing the ellipse arrowheads.

The $r$, $t$ coefficients are given by \cite{Lang_2022_perfect}:

\begin{equation}
t = 1 - \frac{\mathbf{d} \cdot  \mathbf{E}_{f}^* \,\, \mathbf{d}^* \cdot  \mathbf{E}_f}{D} \,,
\label{transmission}
\end{equation}
\begin{equation}
r = -\frac{\mathbf{d} \cdot  \mathbf{E}_{b}^* \,\, \mathbf{d}^* \cdot  \mathbf{E}_f}{D} \,,
\label{reflection}
\end{equation}
with
\begin{equation}
\begin{split}
D = &\frac12 \left( |\mathbf{d} \cdot  \mathbf{E}_{f}^*|^2  + |\mathbf{d} \cdot \mathbf{E}_{b}^*|^2  \right)
+ \zeta \left(\mathbf{d} \cdot  \mathbf{G}_{\text{l}} \cdot  \mathbf{d}^* + \im \hbar\epsilon_0\delta \right)\,.
\end{split}
\end{equation}

We will assume resonance, and thus $\delta$, the detuning between the incident photon and the TLS transition frequency, is set to zero. $\mathbf{G}_{\text{l}}$ is the Green's function controlling the interaction of the TLS with non-guided modes \cite{Hughes_2007_beta}, and the parameter $\zeta$ characterises the inverse of the waveguide mode coupling strength to the TLS. For simplicity we set $\zeta \mathbf{d} \cdot  \mathbf{G}_{\text{l}} \cdot \mathbf{d}^* = L$, with $L$ a scalar controlling loss. If the TLS is prepared in its excited state the fraction of the radiated intensity (photon probability) to enter the guided modes is given by $\beta = W / (W + L)$ with $W = \frac12 ( |\mathbf{d} \cdot  \mathbf{E}_{f}^*|^2  + |\mathbf{d} \cdot \mathbf{E}_{b}^*|^2 )$. By making loss independent of $\di$ we are implicitly assuming at least two loss modes, with polarisations orthogonal at the TLS location.

First, we consider the case of a linear dipole. Here it is well-known that the TLS acts as a mirror, with $|r|$ approaching unity for low loss \cite{Fan_2005_reflection}. However, the phase of the reflection has received less attention.

It should first be clarified what is meant by ``the phase of the reflection''. Naturally this phase must be determined relative to the phase of the input and at some particular location in space. The location is important because the input signal (travelling forwards) will have a phase that changes spatially like $e^{ikx}$, while the backwards travelling reflection will instead evolve as $e^{-ikx}$. As the distance between the phase reference point and the location of the reflecting TLS is increased the phase of the reflection will change according to the additional round trip distance - the mechanism of a Michelson interferometer.

However, if some arbitrary (but fixed) location is picked at which to compare the input and output phases it is then possible to calculate how changing other parameters alters the phase of the reflected light.

\begin{figure}[t]
\includegraphics[scale=0.38]{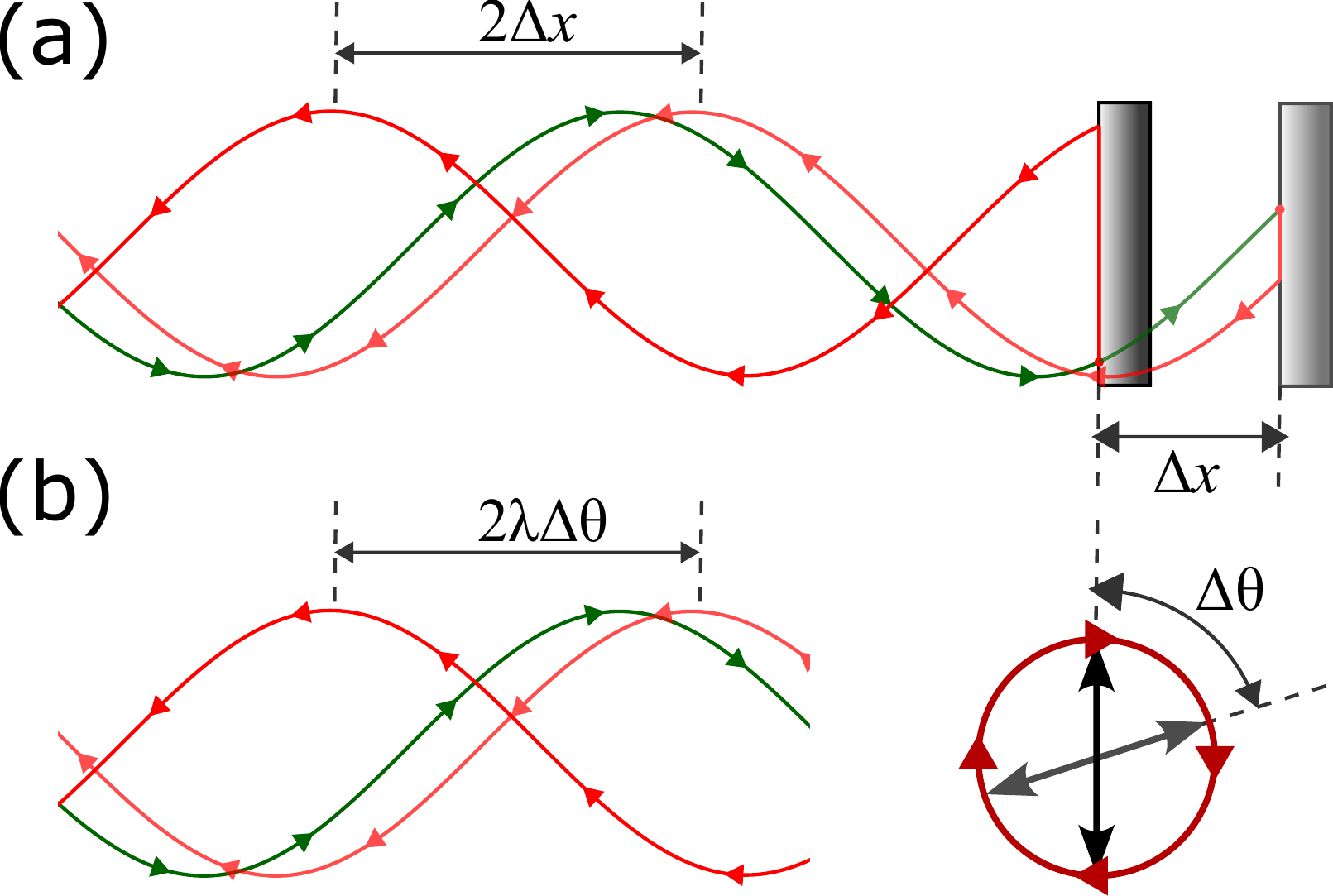}
\caption{(a) Michelson interferometer phase from mirror motion. (b) Comparable effect from rotating a reflecting dipole to interact with a delayed polarisation component.}
\label{interferometer_pic}
\end{figure}

Consider a TLS at a point of circular polarisation, $\Ef = (1, \im)$. Here the $y$ component of the electric field is a quarter-wave delayed relative to the $x$ component. Thus a linear dipole aligned along the $y$ axis is effectively a quarter wavelength  ``further away" than one along $x$. Here rotating a dipole some angle $\Delta \theta$, transforms the reflectivity as $r = r_0 \exp(2 \im \Delta \theta)$, with $r_0$ the reflectivity for $\Delta \theta=0$. The mechanism is reminiscent of a Michelson interferometer, where moving the mirror by distance $\Delta x$ delays the phase by the additional round-trip distance, $2 \Delta x / \lambda$, depicted in fig.\ref{interferometer_pic}. Here we effectively move the mirror by rotating it.

\begin{figure}[t]
\includegraphics[scale=0.13]{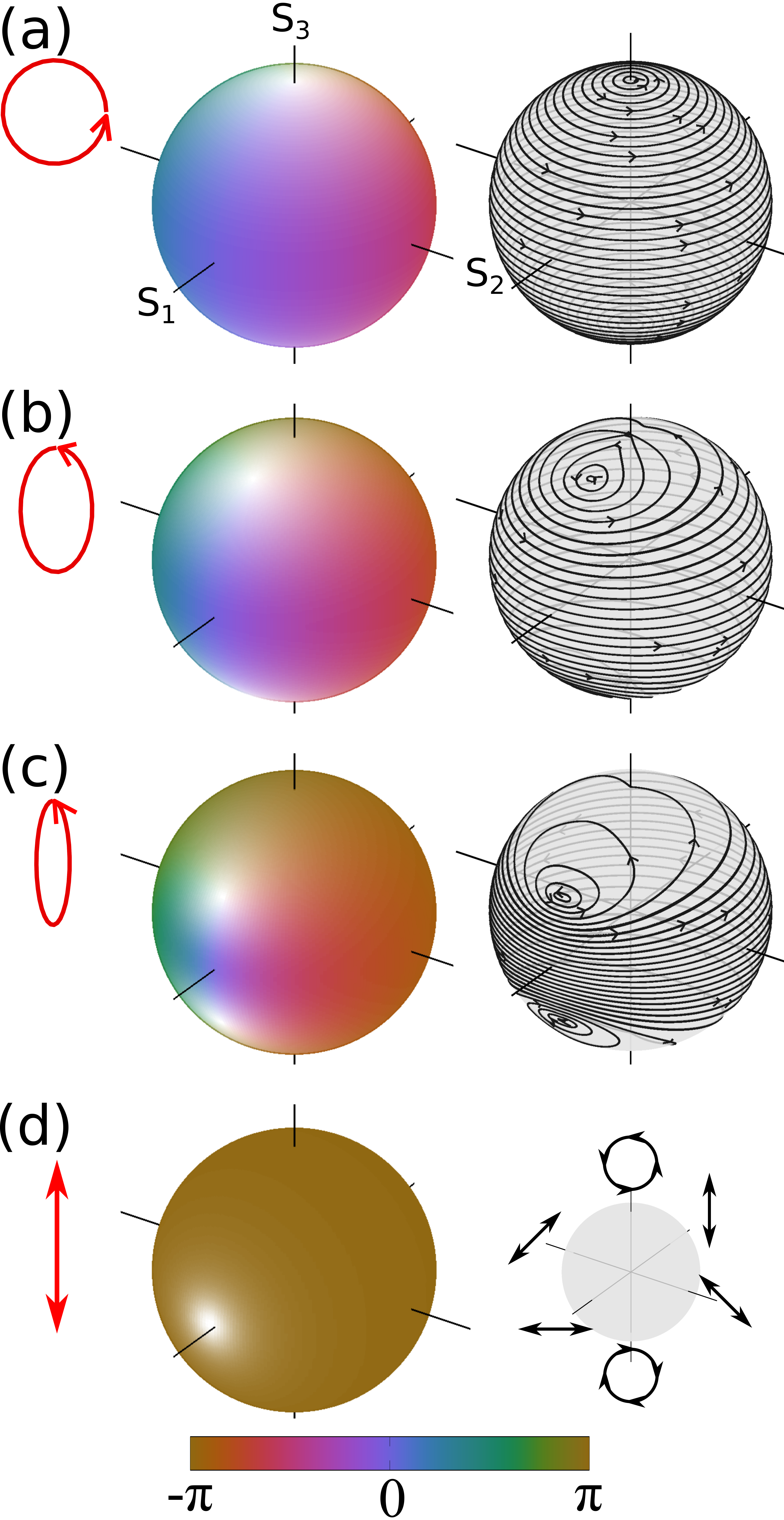}
\caption{$r$ as a function of $\di$. First column: hue (opacity) indicates the phase (amplitude) of the reflection from the dipole on that point of the Poincar{\'e} sphere, for the polarisation depicted on the left. These polarisations are $\Ef = [\im \cos(n\pi/12), \sin(n\pi/12) ]$ with (a,b,c,d) having $n=(3,4,5,6)$ respectively. Second column: streamlines indicating the phase gradient. In (d) the phase is constant so streamlines cannot be plotted. Instead a key indicates the locations of cardinal dipoles on the sphere. $S_{1-3}$ mark the axes of the the Stokes Parameters \cite{Collett_2005_Fieldguide}.}
\label{streamlines}
\end{figure}

This phase is physically meaningful. One could build a interferometer with a rotating dipole replacing the moving mirror and measure the effect. The phase may also be important in other contexts. For example, if multiple TLSs are coupled to a single waveguide their spacing is of critical importance, as it determines whether emission and scattering adds constructively or destructively, thereby controlling the superradiance \cite{Bea_2020_array}, dipole-dipole frequency shifts \cite{Dzsotjan_2011_dipole_dipole, Jones_2018_dipole_surfaces}, reflectivity \cite{Zhou_2020_superrad_reflection} and interatom entanglement \cite{Mirza_2016_multiple_atoms, Zoller_2015_multimer}. But, as we have just motivated, the phase delay is also dependent on the transition dipoles, so that the effective emitter spacing depends on how the TLS dipoles are oriented. Thus, the realisation of proposals that depend on the phase delay between adjacent TLSs  \cite{Holzinger_2022_Dark}, is not determined entirely by the TLS separation.

The ability of both distance and dipole to control phase motivates a mention of superconducting giant atoms. While chirality normally exploits the relative phase between two polarisation components at a single location, these giant atoms achieve a similar effect by reaching spatially to exploit the relative phase between two locations in the waveguide \cite{Nori_2021_giant_atom}.

Requiring that $|\Ef|=|\di|=1$ for simplicity and neglecting global phases both vectors can be written in the form: $[\cos(\theta), \sin(\theta) \exp(i\phi)]$. The angles $\theta$, $\phi$ can be used to represent the vectors as points on the surface of the Poincar{\'e} sphere. In the general case a dipole can be picked from anywhere on the surface of this sphere. The aforementioned linear dipoles comprise the equator, with circular dipoles at the poles and ellipses elsewhere. 

In fig.\ref{streamlines},(a) we fix a circular polarisation, and calculate $r$ for all possible dipoles. Each dipole, $\di$, corresponds to a point on the sphere with its own value of $r$. In the first column of spheres the hue indicates the phase $\angle r$, with the opacity indicating $|r|$ [with $r=|r| \exp(\im \angle r)$]. In the second column the phase gradient is plotted. Following any line in the arrowhead direction the phase changes by $-2\pi$ in a complete cycle. The Michelson-like phase motivated above appears on the equator. We have set $|\Ef|=|\di|=1$ and $L = 0.01$ to consider the situation where coupling to loss modes is weak compared to typical waveguide coupling. Such low loss is appropriate to, for example, photonic crystal waveguide systems \cite{Lorenzo_2019_beta}. Due to this low loss $|r|$ is close to unity almost everywhere on the sphere, becoming significantly less only when the dot products $|\di^* \cdot \Ef|^2$ or $|\di^* \cdot \Eb|^2$ are comparable to $L$.

Plotting the data on a sphere highlights important topological restrictions that are not obvious from inspection of equ.(\ref{transmission}). The phase swirling about the equator directly requires that in each hemisphere there is a dipole such that $r = 0$. This can be seen from the figure, after fixing the equator one cannot find a smooth function without at least one zero in each hemisphere.

The zero points are \emph{phase singularities}, a ubiquitous wave phenomenon that occur where a complex scalar field takes value $0$ at some location, with the phase angle varying by $2\pi m$ in a circuit of the zero point \cite{Nye_1974_wave_trains, Berry_2000_tides}. The total phase change along a closed curve is equal to the $\sum_n 2 \pi m_n$ with $n$ counting over the singularities enclosed. Thus, our $2\pi$ variation along the equator is enough to ensure that both hemispheres contain at least one phase singularity with $r = 0$.

With circular polarisation these points are the poles. At one $\di^* \cdot \Ef = 0$ and the dipole decouples from the forwards mode, so the TLS cannot in any way interact with the input photon. At the other $\di^*\cdot \Eb = 0$ so that whatever the TLS does to the photon it cannot involve any scatting to the backward mode (hence $r=0$).

Moving to parts (b, c, d) of the figure we vary the polarisation of the waveguide. This deforms the phase structures continuously, preserving the two singularities until they mutually annihilate on the equator for a linear polarisation. This highlights that for any polarisation there is a dipole that decouples from the forward mode and another the backward one (the singularities), with the two coinciding only for a linear polarisation \cite{Lang_2022_perfect}.

The phase gradient can be considered a vector field. As this field lives on the sphere it is subject to the ``hairy ball theorem" which requires that it cannot be smooth and nonzero everywhere. The theorem name refers to a consequence of this, that a hairy ball cannot be combed to have all the hair lie flat. More precisely the theorem requires the total Poincar{\'e}-Hopf indices of the vector field's zero points equal the sphere's Euler characteristic of +2 \cite{Berry_2004_sky}. The vector fields depicted by streamlines in fig.\ref{streamlines} each have two singular points where the vector winds in a circle (the phase singularities), such circles have an index of +1 irrespective of the arrowhead directions, so that the pair has the required +2 total.

\begin{figure}[t]
\includegraphics[scale=0.36]{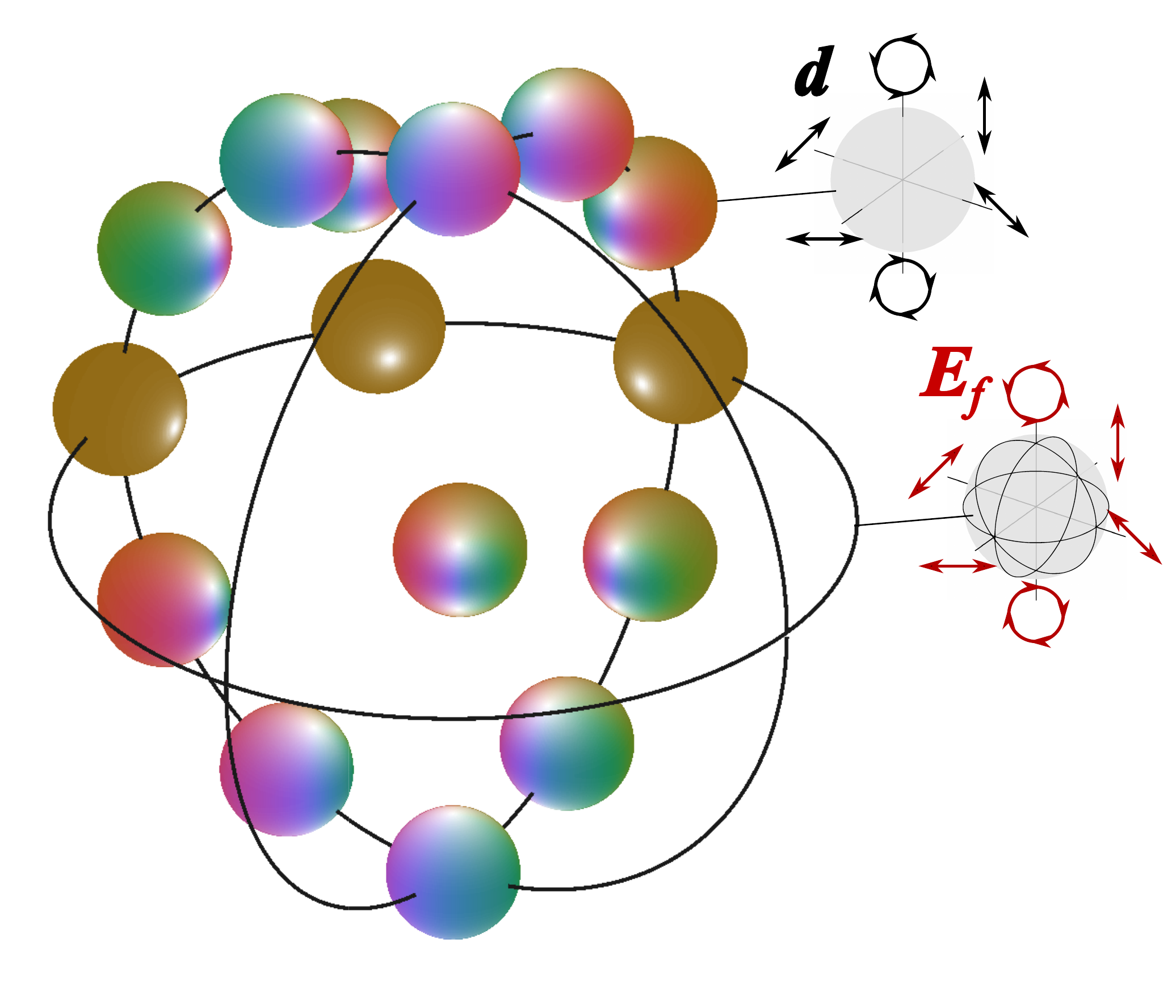}
\caption{Each small sphere denotes $r$ as a function of $\di$ as in fig.\ref{streamlines}. For each the polarisation, $\Ef$ used is indicated by its location on the larger (wire frame) sphere.}
\label{sphere_of_spheres}
\end{figure}

The polarisation is able to explore its own Poincar{\'e} sphere of possible values, such that the space of $\Ef \otimes \mathbf{d}$ has the form $\mathbb{S}^2 \otimes \mathbb{S}^2$, a  ``sphere of spheres" which we term a Poincar{\'e} hypersphere. This full space is depicted in fig.\ref{sphere_of_spheres}, with the larger sphere indicating the polarisation and the smaller ones the dipole. One sees that (starting from the north pole of the larger sphere) stretching our initially circular polarisation brings the polar phase singularities closer to one another, and that the line of latitude moved along is determined by the latitude line on the big sphere. Opposite points on the Poincar{\'e} sphere correspond to orthogonal polarisations, so that one phase singularity on each sub-sphere points to the centre of the larger sphere. This corresponds to the $\di^* \cdot \Ef =0$ case. The other phase singularity's location on each sub-sphere is given by reflecting the first through the equator to set $\di^* \cdot \Eb =0$.

Taking a tangent, it is interesting to consider the geometry. $\Ef$ and $\di$ are each 2D, giving us a total of 4 dimensions. A phase singularity has 2 dimensions fewer than the space it is embedded in, so that the singularities on our 2-spheres were pointlike (0D), and in 3D space singularities represent lines of darkness or silence in fields \cite{Denis_2009_threads_of_darkness}. Our singularities in the 4D space are 2D surfaces. There are two such surfaces, each corresponding to $\di^* \cdot (\Ef \text{ or } \Eb)=0$. These surfaces map to spherical shells, and the intersection of the two maps to a circle. On this circle both dipole and polarisation are linear and are orthongonal to one another, for example resembling a ``$\times$" in real space. The circular nature of the intersection relates to the fact that the ``$\times$" can be rotated freely, eg. ``$+$".

The transmission, $t$ can be equally assessed on the sphere (or hypersphere). However $t$ lacks the complex structure seen in $r$. For the most part, a $t$ equivalent of fig.\ref{streamlines} simply shows a phase of $\angle t = \pi$ in one hemisphere and $\angle t =0$ in the other, the two separated by a $|t|=0$ equator. This $|t|=0$ equator can be considered a phase singularity by re-introducing the detuning, $\delta$. For a fixed $\Ef$, within the 3D space defined by $\di \otimes \delta$ it takes the form of a 1D line phase singularity, looped into a circle (near the equator) in the $\delta = 0$ slice.

In conclusion, we demonstrate a rich behaviour of the reflection coefficient, $r$, of a complex dipole TLS in a one-dimensional photonic waveguide. $r$ is a complex field supporting phase singularities that live on the Poincar{\'e} (hyper)sphere. The existence of two dipoles for each polarisation such that $r=0$ and the existence of a Michelson-interferometer like phase gradient on the equator can both be motivated by physical arguments. We showed that the two effects are intimately linked, due to the topological constraints faced by phase singularities living on the surface of a sphere. The rich dependence of $r$ on dipole and waveguide polarisation for a single TLS will offer new avenues for exploitation of chains of TLSs in waveguides.

{\it{Acknowledgement}}: We acknowledge support from EPSRC grants no. EP/N003381/1 ``One-dimensional quantum emitters and photons for quantum technologies" and EP/M024156/1 ``Spin Space" and PHD funding DTA-1407622.






\bibliography{references}

\end{document}